\documentclass[a4paper,final,conference]{IEEEtran}
\usepackage{mathpple}
\usepackage{times}

\usepackage[top=0.6in, bottom=0.7in, left=0.6in, right=0.6in]{geometry}

\usepackage{amsmath}  
\usepackage{amssymb}  
\usepackage{mathrsfs} 

\usepackage{theorem}  
\usepackage{cite}     
\usepackage{comment}  

\usepackage{upref}
\usepackage{amsfonts}

\usepackage{verbatim}

\usepackage[dvipsnames,usenames]{color}
\usepackage{enumerate}

\usepackage{graphicx}

\usepackage{latexsym}

\usepackage{color}

\usepackage{multicol}
\usepackage{hhline}

\usepackage{psfrag}

\usepackage{tikz}
\usepackage{pgfplots}

\usepackage{hhline}

\usepackage{algorithm}
\usepackage[noend]{algpseudocode}

\usepackage{xcolor,colortbl}

\algnewcommand{\Initialize}[1]{%
	\State \textbf{Initialize:}
	\Statex \hspace*{\algorithmicindent}\parbox[t]{.8\linewidth}{\raggedright #1}
}

\newcommand{\be}[1]{\begin{equation}\label{#1}}
\newcommand{\ee}{\end{equation}}

\newcommand{\bc}{\begin{center}}
\newcommand{\ec}{\end{center}}

\newcommand{\floor}[1]{\lfloor{#1}\rfloor}

\newcommand{\qed}{\hfill$\Box$\\[1ex]}

\newcommand{\cC}{{\cal C}}

\newcommand{\cO}{{\cal O}}

\newcommand{\bfi}{{\boldsymbol i}}

\newcommand{\bfv}{{\boldsymbol v}}
\newcommand{\bfw}{{\boldsymbol w}}
\newcommand{\bfx}{{\boldsymbol x}}
\newcommand{\bfy}{{\boldsymbol y}}
\newcommand{\bfz}{{\boldsymbol z}}

\renewcommand{\leq}{\leqslant}

\renewcommand{\geq}{\geqslant}

\newcommand{\deff}{\mbox{$\stackrel{\rm def}{=}$}}

\newcommand{\F}{\mathbb{F}}

\newcommand{\Cref}[1]{Co\-rol\-la\-ry\,\ref{#1}}

\theoremstyle{plain} \theorembodyfont{\normalfont\slshape}

\newtheorem{thm}{Theorem$\!$}
\newenvironment{theorem}{\begin{thm}\hspace*{-1ex}{\bf.}}{\end{thm}}

\newtheorem{prop}[thm]{Proposition$\!$}

\newtheorem{lem}[thm]{Lemma$\!$}
\newenvironment{lemma}{\begin{lem}\hspace*{-1ex}{\bf.}}{\end{lem}}

\newtheorem{cor}[thm]{Corollary$\!$}
\newenvironment{corollary}{\begin{cor}\hspace*{-1ex}{\bf.}}{\end{cor}}

\newtheorem{prob}[thm]{Problem$\!$}

\newtheorem{defi}[thm]{Definition$\!$}
\newenvironment{definition}{\begin{defi}\hspace*{-1ex}{\bf.}}{\end{defi}}
\newenvironment{Definition}{\begin{defi}\hspace*{-1ex}{\bf.}}{\end{defi}}

\newtheorem{cons}{Construction}

\theorembodyfont{\normalfont}

\newtheorem{exam}{Example$\!$}

\newtheorem{remrk}{Remark$\!$}

\definecolor{Codecolor}{named}{White}  

\newcommand{\Copen}{\mbox{\{\kern-5.50pt\{}}
\newcommand{\Cclose}{\mbox{\}\kern-5.50pt\}}}
\newcommand{\Cslash}{\mbox{$\backslash\kern-6.02pt\backslash$}}

\begin{document}

\title{\textbf{Nearly Optimal Constructions of PIR and \\Batch Codes}\vspace{-1.5ex}}

\date{\today}
\author{
\IEEEauthorblockN{\textbf{Hilal Asi}}
\IEEEauthorblockA{
Technion - Israel Institute of Technology \\
Haifa 32000, Israel \\
{\it shelal@cs.technion.ac.il}\vspace*{-6.3ex}}
\and
\IEEEauthorblockN{\textbf{Eitan Yaakobi}}
\IEEEauthorblockA{
Technion - Israel Institute of Technology\\
Haifa 32000, Israel \\
{\it yaakobi@cs.technion.ac.il}\vspace*{-6.3ex}}
}
\maketitle

\thispagestyle{empty}

\begin{abstract}
In this work we study two families of codes with availability, namely \emph{private information retrieval (PIR) codes} and \emph{batch codes}. While the former requires that every information symbol has $k$ mutually disjoint recovering sets, the latter asks this property for every multiset request of $k$ information symbols. The main problem under this paradigm is to minimize the number of redundancy symbols. We denote this value by $r_P(n,k), r_B(n,k)$, for PIR, batch codes, respectively, where $n$ is the number of information symbols. Previous results showed that for any constant $k$, $r_P(n,k) = \Theta(\sqrt{n})$ and $r_B(n,k)=\cO(\sqrt{n}\log(n))$. In this work we study the asymptotic behavior of these codes for non-constant $k$ and specifically for $k=\Theta(n^\epsilon)$. We also study the largest value of $k$ such that the rate of the codes approaches 1, and show that for all $\epsilon<1$, $r_P(n,n^\epsilon) = o(n)$, while for batch codes, this property holds for all $\epsilon< 0.5$.
\vspace{-1ex}
\end{abstract}

\section{Introduction}\label{sec:intro}
\renewcommand{\baselinestretch}{0.99}\normalsize
In this paper we study two families of codes with availability for distributed storage. The first family of codes, called \emph{private information retrieval} (\emph{PIR}) \emph{Codes}, requires that every information symbol has some $k$ mutually disjoint recovering sets. These codes were studied recently in~\cite{FVY15} due to their applicability for private information retrieval in a coded storage system. They are also very similar to \emph{one-step majority-logic decodable codes} that were studied a while ago by Massey~\cite{Massey} and later by Lin and others~\cite{LC04} and were prompted by applications of error-correction with low-complexity.

The second family of codes, which is a generalization of the first one, was first proposed in the last decade by Ishai et al. under the framework of \emph{batch codes}~\cite{IKOS04}. These codes were originally motivated by different applications such as load-balancing in storage and cryptographic protocols. Here it is required that every multiset request of $k$ symbols can be recovered by $k$ mutually disjoint recovering sets. 

Formally, we denote a $k$-PIR code by $[N,n,k]^P$ to be a coding scheme which encodes $n$ information bits to $N$ bits such that each information bit has $k$ mutually disjoint recovering sets. Similarly, a $k$-batch code will be denoted by $[N,n,k]^B$ and the requirement of mutually disjoint recovering sets is imposed for every multiset request of size $k$.  The main figure of merit when studying PIR and batch codes is the value of $N$, given $n$ and $k$. Thus, we denote by $P(n,k), B(n,k)$ the minimum value of $N$ for which an $[N,n,k]^P, [N,n,k]^B$ code exists, respectively. 

Since it is known that for all fixed $k$, $\lim_{n\rightarrow \infty} B_q(n,k)/n=\lim_{n\rightarrow \infty} P_q(n,k)/n=1$,~\cite{IKOS04}, we evaluate these codes by their redundancy and define $r_B(n,k) \triangleq B(n,k)-n, r_P(n,k) \triangleq P(n,k)-n$. One of the problems we study in the paper studies the largest value of $k$ (as a function of $n$) for which one can still have $r_P(n,k)=o(n)$ and $r_B(n,k)=o(n)$, so the rate of the codes approaches 1. We show that for PIR codes this holds for $k=\Theta(n^\epsilon)$, for all $\epsilon<1$, while for batch codes for all $\epsilon<1/2$. Since $r_P(n,k), r_B(n,k)\geq k$, the result for PIR codes is indeed optimal. Furthermore, in order to have a better understanding of the asymptotic behavior of the redundancy, we study the values $r_P(n,k)$ and $r_B(n,k)$ when $k=\Theta(n^\epsilon)$. 

The results we achieve in the paper are based on two constructions. The first one uses \emph{multiplicity codes} which generalized Reed Muller codes and were first presented by Kopparty et al. in~\cite{KSY10}. These codes were also used for the construction of \emph{locally decodable codes}~\cite{Y12}. The second construction we use is based on the subcube construction from~\cite{IKOS04}. This basic construction can be used to construct both PIR and batch codes. While the idea in the works in~\cite{IKOS04,FVY15} was to use multidimensional cubes in order to achieve large values of $k$, here we take a different approach and position the information bits in a two dimensional array and then form multiple parity sets by taking different diagonals in the array.

The rest of the paper is organized as follows. In Section~\ref{sec:def}, we formally define the codes studied in this paper and review previous results. In Section~\ref{sec:multiplicity}, we review multiplicity codes. Then, in Section~\ref{sec:multiPIR} we show how to use multiplicity codes to construct PIR codes, and in Section~\ref{sec:multi_batch} we carry the same task for batch codes. 
Then, in Section~\ref{sec:arrayPIR}, we present our array construction and its results for PIR codes and batch codes. Due to the lack of space some proofs in the paper are omitted.

\vspace{-1ex}
\section{Definitions and Preliminaries}\label{sec:def}

Let $\F_q$ denote the field of size $q$, where $q$ is a prime power. A linear code of length $N$ and dimension $n$ over $\F_q$ will be denoted by $[N,n]_q$. 
For binary codes we will remove the notation of the field. The set $[n]$ denotes the set of integers $\{1,2,\ldots,n\}$.

In this work we focus on two families of codes, namely \emph{private information retrieval} (\emph{PIR}) codes that were defined recently in~\cite{FVY15} and \emph{batch codes} that were first studied by Ishai et al. in~\cite{IKOS04}. Formally, these codes are defined as follows. \vspace{-1.5ex}
\begin{definition}
Let $\cC$ be an $[N,n]_q$ linear code over the field $\F_q$. 
\begin{enumerate}
\item The code $\cC$ will be called a \textbf{$k$-PIR code}, and will be denoted by $[N,n,k]_q^P$, if for every information symbol $x_i, i \in [n]$, there exist $k$ mutually disjoint sets $R_{i,0},\ldots, R_{i,k-1}\subseteq [N]$ such that for all $j\in [k]$, $x_i$ is a function of the symbols in $R_{i,j}$.
\item The code $\cC$ will be called a \textbf{$k$-batch code}, and will be denoted by $[N,n,k]_q^B$, if for every multiset request of symbols 
$\{i_0,i_1,\ldots,i_{k-1}\}$, there exist $k$ mutually disjoint sets $R_{i_0},R_{i_1},\ldots, R_{i_{k-1}}\subseteq [N]$ such that for all $j\in [k]$, $x_{i_j}$ is a function of the symbols in $R_{i_j}$.\vspace{-1ex}
\end{enumerate}
\end{definition}

We slightly modified here the definition of batch codes. In their conventional definition, $n$ symbols are encoded into some $m$ tuples of strings, called buckets, such that each batch (i.e. request) of $k$ information symbols can be decoded by reading at most some $t$ symbols from each bucket. In case each bucket can store a single symbol, these codes are called \emph{primitive batch codes}, which is the setup we study here and for simplicity call them batch codes. In this work we study the binary and non-binary cases of PIR and batch codes.

The main problem in studying PIR and batch codes is to minimize the length $N$ given the values of $n$ and $k$. We denote by $P_q(n,k),B_q(n,k)$ the value of the smallest $N$ such that there exists an $[N,n,k]_q^P,[N,n,k]_q^B$ code, respectively. Since every batch code is also a PIR code with the same parameters we get that $B_q(n,k) \geq P_q(n,k)$. For the binary case, we will remove $q$ from these and subsequent notations.

In~\cite{IKOS04}, it was shown using the subcube construction that for any fixed $k$ there exists an asymptotically optimal construction of $[N,n,k]_q^B$ batch code, and hence\vspace{-1ex} $$\lim_{n\rightarrow \infty} B_q(n,k)/n=\lim_{n\rightarrow \infty} P_q(n,k)/n=1.\vspace{-1ex}$$ Therefore, it is important to study how fast the rate of these codes converges to one, and so the redundancy of PIR and batch codes is studied. We define $r_B(n,k)_q$ to be the value $r_B(n,k)_q \triangleq B(n,k)_q-n$ and similarly, $r_P(n,k)_q \triangleq P(n,k)_q-n$. 

In~\cite{FVY15}, it was shown that for any fixed $k\geq3$ there exists an $[N,n,k]$ PIR code where $N=n+\cO(\sqrt{n})$, so $r_P(n,3)=\cO(\sqrt{n})$ and in~\cite{RV16} it was proved that $r_P(n,3)=\Theta(\sqrt{n})$, by providing a lower bound on the redundancy of 3-PIR codes. These results assure also that for any fixed $k$, $r_P(n,k)=\Theta(\sqrt{n})$ and also implied that for any fixed $k$, $r_B(n,k)=\Omega(\sqrt{n})$. In~\cite{VY16}, it was proved that for $k=3,4$, $r_B(n,k)=\Theta(\sqrt{n})$, and for any fixed $k\geq 5$, $r_B(n,k)=\cO(\sqrt{n}\log(n))$.
In this paper, we will mostly study the values of $r_P(n,k)$ and $r_B(n,k)$, when $k$ is a function of $n$, for example $k=\Theta(n^\epsilon)$. One of the problems we will also investigate is finding the largest $\epsilon$ for which $r_P\big(n,k=\Theta(n^\epsilon)\big)=o(n)$, and similarly for batch codes.

There are several more constructions of PIR and batch codes, which we summarize below.
\begin{enumerate}
\item $r_B(n,n^{1/3})\leq n$,~\cite{RSDG16}.
\item $r_B(n,n^{\epsilon})\leq n^{7/8}$ for $7/32\leq \epsilon \leq 1/4$,~\cite{RSDG16}.
\item $r_B(n,n^{\epsilon})\leq n^{4\epsilon}$ for $1/5<\epsilon\leq 7/32$,~\cite{RSDG16}.
\item $B(n,n) \leq 2n^{1.5}$,~\cite{BYCS16}.
\item $r_P(n,\sqrt{n})= O(n^{(\log3)/{2}})$,~\cite{LC04}.
\item $r_P(n,n^{\epsilon})= O(n^{0.5+\epsilon})$,~\cite{LC04}.
\end{enumerate}
\vspace{-1ex}

\section{Mutliplicity Codes}\label{sec:multiplicity}
In this section we review the construction of \emph{multiplicity codes}. This family of codes was first presented by Kopparty et al. in~\cite{KSY10} as a generalization of Reed Muller codes by calculating the derivatives of polynomials. We follow the definitions of these codes as were presented in~\cite{KSY10} and first start with the definition of the Hasse derivative.

For a field $\F$, let $\F[x_1,\dots,x_s]=\F[\bfx]$ be the ring of polynomials in the variables $x_1,\dots,x_s$ with coefficients in $\F$. For a vector $\bfi=(i_1,\dots,i_s)$ of non-negative integers, its weight $wt(\bfi)$ is $\sum_{j=1}^{s} i_j$, and let $\bfx^\bfi$ denote the monomial $\prod_{j=1}^{s} x_j^{i_j}$. The total degree of this monomial equals $wt(\bfi)$. For $P(\bfx)\in{\F[\bfx]}$, let the degree of $P(\bfx)$, $\deg(P)$, be the maximum total degree over all monomials in $P(\bfx)$. \vspace{-1ex}

\begin{Definition}
	For a polynomial $P(\bfx)\in{\F[\bfx]}$ and a non-negative vector $\bfi$, the $\bfi$-th \textbf{Hasse derivative} of $P(\bfx)$, denoted by $P^{(\bfi)}(\bfx)$, is the coefficient of $\bfz^{\bfi}$ in the polynomial $P'(\bfx,\bfz)=P(\bfx+\bfz)\in{\F[\bfx,\bfz]}$.
\end{Definition}

\begin{definition}
	Let $m,d,s$ be nonnegative integers and let $q$ be a prime power. Let
	$ \Sigma = \F_q^{ |\{\bfi : wt(\bfi)<m \}|}=  \F_q^{ {{s+m-1} \choose s} } .$
	For a polynomial $P(x_1,\dots,x_s)\in{\F_q[x_1,\dots,x_s]}$, we define the order $m$ evaluation of $P$ at $\bfw\in \F_q^s$, denoted by $P^{(<m)}(\bfw)$, to be the vector\vspace{-1ex}
	$$ P^{(<m)}(\bfw) = \big(  P^{(\bfi)}(\bfw)  \big)_{\bfi:wt(\bfi)<m}\in{\Sigma}.\vspace{-1ex}$$
	The \textbf{multiplicity code }$\cC(m,d,s,q)$ of order $m$ evaluations of degree $d$ polynomials in $s$ variables is defined as follows. The code is over $\Sigma$, has length $q^s$, and its coordinates are indexed by elements in $\F_q^s$. For each polynomial $P(\bfx)\in{\F_q[x_1,\dots,x_s]}$ with $\deg(P)\leq{d}$, there is a codeword in $\cC$ given by:
	$Enc_{m,d,s,q}(P) =  \big(  P^{(<m)}(\bfw)  \big)_{\bfw\in{\F_q^s}}\in{(\Sigma})^{q^s}.$
	That is,\vspace{-1ex}
	$$\hspace{-0.3ex}\cC(m,d,s,q) \hspace{-0.5ex}=\hspace{-0.5ex}  \{ \hspace{-0.3ex} \big(  P^{(<m)}\hspace{-0.3ex}(\bfw)  \big)_{\bfw\in{\F_q^s}}\hspace{-0.5ex}\in\hspace{-0.5ex}{\Sigma}^{q^s} \hspace{-1ex}:\hspace{-0.3ex} P \hspace{-0.5ex}\in\hspace{-0.5ex} \F_q[\bfx], \deg(P)\hspace{-0.5ex}\leq \hspace{-0.5ex}d \}.\vspace{-1ex}$$
\end{definition}

The following lemma was proved in~\cite{KSY10}, Lemma 9.\vspace{-2ex}
\begin{lemma}\label{mult_dis_rate}
	The multiplicity code $\cC(m,d,s,q)$ has relative  distance at least $\delta = 1 - \frac d {mq}$ and rate $ { { {d+s} \choose {s}  }}  /{  { {s+m-1} \choose {s}  }q^s  } $.
\end{lemma}\vspace{-1ex}
Lastly, we note that since the multiplicity code $\cC(m,d,s,q)$ is a linear code it can also be a systematic code and thus for the rest of the paper we assume these codes to be systematic; for more details see Lemma 2.3 in~\cite{Y12}. For the rest of the paper and unless stated otherwise, we assume that $m,d,s,q$ are positive integers. \vspace{-0.2ex}

\section{PIR Codes from multiplicity codes}\label{sec:multiPIR}
In~\cite{KSY10}, multiplicity codes were used to construct \emph{locally decodable codes} in order to retrieve the value of a single symbol with high probability, given that at most a fixed fraction of the codeword's symbol has errors~\cite{Y12}. Since we are not concerned with errors, we modify the recovering procedure so that each information symbol has a large number of disjoint recovering sets. For this end, we establish several properties on interpolation sets of polynomials which will help us later to construct the recovering sets, and thus PIR and batch codes.
\vspace{-4ex}
\begin{lemma}\label{homogeneous_IS}
	Let $P(\bfx)\in{\F_q[x_1,\dots,x_s]}$ be an homogeneous polynomial\footnote{We say that $P(\bfx)\in{\F_q[\bfx]}$ is homogeneous if all the monomials of $P(\bfx)$ have the same total degree.} such that $\deg(P)=d$. Let $A_1,\dots,A_{s-1}$ be subsets of $\F_q$ such that $|A_i|=d+1$. Then the set $A=A_1 \times \dots \times A_{s-1}\times \{1\}$ is an interpolation set\footnote{For $P(\bfx)\in{\F_q[x_1,\dots,x_s]}$ and $R\subseteq \F^s_q$, we say that $R$ is an interpolation set of $P(\bfx)$ if for every polynomial $Q(\bfx)$ such that $P(\bfx)=Q(\bfx)$ for every $\bfx\in R$, it holds that $P(\bfx)=Q(\bfx)$ for every $\bfx\in \F_q^s$.} of $P(\bfx)$, where $1\in{\F_q}$ is the unitary element of the field.
\end{lemma}\vspace{-1ex}

The following definition will be used in the construction of recovering sets for multiplicity codes.\vspace{-2ex}
\begin{definition}
	Let $\F_q$ be a field, and $S_1,S_2\subseteq{\F_q^s}$ where $s$ is a positive integer. We say that the sets $S_1$ and $S_2$ are \textbf{disjoint under multiplication} if for every $x\in{S_1}$ and $\alpha \in{\F_q\setminus\{0\}}$ it holds that $\alpha x \notin {S_2}$.\vspace{-1ex}
\end{definition}\vspace{-2ex}

\begin{lemma}\label{lemma_homogeneous_disjoint_IS}
	Let $P(\bfx)\in{\F_q[x_1,\dots,x_s]}$ be an homogeneous polynomial such that $\deg(P)=d$. Then there exists $\floor{\frac q {d+1}}^{s-1}$ interpolation sets of $P(\bfx)$, each of size $(d+1)^{s-1}$, which are mutually disjoint under multiplication.
\end{lemma}\vspace{-1ex}

Now we are in a good position to present the recovering procedure for multiplicity codes. First, we show a general structure of the recovering sets, and then we argue that many disjoint sets can be constructed this way.\vspace{-2ex}
\begin{theorem}\label{MULT_recovering_sets}
	Let $m,d,s,q$ be such that $d/m < q-1$, and $\cC=\cC(m,d,s,q)$ is the multiplicity code of length $q^s$ over  $\F_q^{ {{s+m-1} \choose s} }$. Let $A\subseteq \F_q^s$ be an interpolation set for homogeneous polynomials of degree at most $m-1$. Then, for every $\bfy=(y_{\bfw})_{\bfw\in \F_q^s}\in \cC$, and for any $\bfw_0\in \F_q^s$, the set of coordinates indexed by the set\vspace{-1ex}
	$$R=\{ \bfw_0 \}+\F_qA \triangleq \{ \bfw_0  +\lambda \bfv : \bfv\in{A}, \lambda\in{\F_q}\setminus \{0\} \}\vspace{-1ex}$$ is a recovering set for the symbol $y_{\bfw_0}$.  \vspace{-1ex}
\end{theorem}

\begin{IEEEproof}
	The proof follows similar ideas to the one from~\cite{KSY10}. 
	Recall that every codeword $\bfy=(y_{\bfw})_{\bfw\in \F_q^s}\in \cC$ corresponds to a polynomial $P(\bfx)\in \F_q[\bfx]$, of degree at most $d$, where for all $\bfw\in \F_q^s$, $y_\bfw = P^{(<m)}(\bfw)$. Every vector $\bfv$ in the interpolation set $A$ is called a \emph{direction} and will correspond to a line containing $\bfw_0$ in the direction $\bfv$. Reading the order $m$ evaluations of the polynomial $P(\bfx)$ at these lines will enable us to recover the value of $P^{(<m)}(\bfw_0)$. This procedure consists of two steps, described as follows.\\
	\textbf{Step 1:}	For every direction $\bfv\in{A}$, define the following univariate polynomial 
$	p_\bfv(\lambda)=P(\bfw_0+\lambda \bfv)\deff {\sum_{j=0}^{d} c_{\bfv,j}\lambda^j} \in \F_q[x]$.
	Since the values and the derivatives of $P(\bfw_0+\lambda \bfv)$ for all $\lambda \in{\F_q} \setminus{\{0\}}$ are known, and $\deg(p_\bfv)\leq{d}$, one can prove, as in~\cite{KSY10}, that $p_\bfv(\lambda)$ is unique, and thus can be recovered.\\
	\textbf{Step 2}:
	From Step 1, one can get that \vspace{-1ex}
	$$p_\bfv(\lambda) = \sum_{\bfi} P^{(\bfi)}(\bfw_0)\bfv^\bfi\lambda^{wt(\bfi)}={\sum_{j=0}^{d} c_{\bfv,j}\lambda^j},\vspace{-1ex}$$
	and therefore for  $0\leq j\leq d$,
	$\sum_{\bfi : wt(\bfi)=j} P^{(\bfi)}(\bfw_0)\bfv^\bfi = c_{\bfv,j}.$
	{Considering only the first $m$ of these $d+1$ equations, we get that} $u_\bfi=P^{(\bfi)}(\bfw_0)$ is a solution for the equations system
	\begin{equation}\label{linear_system}
	\sum_{\bfi : wt(\bfi)=j} u_\bfi \bfv^\bfi = c_{\bfv,j} , \ \ 0\leq{j}<m\leq d.\vspace{-1.5ex}
	\end{equation}
	Now we prove that the equations system~(\ref{linear_system}) has a unique solution. Indeed, if we denote $Q_j(\bfx)=\sum_{\bfi:wt(\bfi)=j} u_\bfi\bfx^\bfi \in\F_q[x_1,\ldots,x_s]$ where $0\leq{j}<m$, we get that the equations in~(\ref{linear_system}) are equivalent to
	$Q_j(\bfv) = c_{\bfv,j}$
	for every $\bfv\in{A}$. But since for every $j$ we know that $Q_j$ is an homogeneous polynomial of degree $j$, and $A$ is an interpolation set for homogeneous polynomials of degree at most $m-1$, we get that the polynomial $Q_j(\bfx)$ is unique. Therefore, we can recover the value of $P^{(<m)}(\bfw_0)$ by solving the equations system~(\ref{linear_system}).
\end{IEEEproof}

The next theorem shows how to construct PIR codes from Multiplicity Codes. \vspace{-2ex}
\begin{theorem}\label{PIR_from_mult}
	For all $m,d,s,q$ such that $\frac d m < q-1$, the code $\cC(m,d,s,q)$ is a $k$-PIR code $[q^s,n,k]_Q^P$, where $n=\frac { { {d+s} \choose {s}  }}  {  { {s+m-1} \choose {s}  }  }, k=\floor{\frac q {m}}^{s-1}$, and $Q= q^{{s+m-1} \choose {s}}$. 
\end{theorem}\vspace{-1ex}
\begin{IEEEproof}
	According to Theorem~\ref{MULT_recovering_sets}, every interpolation set $A$ for homogeneous polynomials of degree $m-1$ defines a recovering set, which consists of the lines containing $\bfw_0$ in the directions of $\bfv$ for all $\bfv\in{A}$. Therefore, in order to get disjoint recovering sets, all we need to do is to pick different lines. According to Lemma~\ref{lemma_homogeneous_disjoint_IS}, there are $\floor{\frac q {m}}^{s-1}$ interpolation sets for homogeneous polynomials of degree $m-1$ which are mutually disjoint under multiplication. This means that each line cannot appear in two sets, thus the recovering sets defined by these interpolation sets are disjoint.
\end{IEEEproof}

The next theorem summarizes the results in this section.\vspace{-2ex}
\begin{theorem}\label{theorem_PIR_summary}
	For every positive integer $s\geq 2$, $0<\alpha <  {1}$, and $n$ sufficiently large, there exists a $k$-PIR code $[N,n,k]_Q^P$, over $\F_Q$ with redundancy $r=N-n$ such that\vspace{-1ex}
	\begin{align*}
	k = \Theta(n^{(1-{\frac 1 s})(1-\alpha)}), Q = n^{\Theta(n^\alpha)}, r = \cO(n^{1-\frac \alpha s}).	
	\end{align*} 
In particular, for $0 \leq \epsilon <1$, it holds that $r_P\big(n,k=\Theta(n^\epsilon)\big) = \cO\big(n^{\delta(\epsilon)}\big),$ where 
$\delta(\epsilon) = \min_{{s:s > {\frac 1 {1-\epsilon}}}} \{\delta_s(\epsilon)\},$ and  $\delta_s(\epsilon)= 1-\frac 1 {s}+\frac \epsilon {s-1}$. For a given value of $\epsilon$, the value $s^*$ that minimizes $\delta(\epsilon)$ is
$s^* =\floor{  \frac{2}{1-\epsilon}  }$.
\end{theorem}
\vspace{-1ex}

Now we use our last result in order to construct binary $k$-PIR codes. The main idea is to convert every symbol of the field $\F_Q$ to $\log(Q)$ binary symbols. We say that $f(n)=\Omega(n^{a^-})$ is for all $\tau>0$, $f(n)=\Omega(n^{a-\tau})$. Similarly we define $f(n)=\cO(n^{a^+})$ if for all $\tau>0$, $f(n)=\cO(n^{a+\tau})$.\vspace{-2ex}
\begin{theorem}\label{PIR_binary_summary}
For every positive integer $2\leq{s}$, $0<\alpha <1$, and $n$ sufficiently large, there exists a binary $k$-PIR code $[N,n,k]^P$, with redundancy $r=N-n$ such that\vspace{-1ex}
\begin{align*}
k = {\Theta\big(\big(\frac{n}{\log(n)}\big)^{(1-{\frac 1 s}) \frac{ 1-\alpha} {1+\alpha}  }\big)}, r = \cO\big(n^{ 1-\frac {\alpha} {s(1+\alpha)}  } (\log(n))^{\frac{\alpha}{s(1+\alpha)}}\big).\vspace{-3ex}
\end{align*}
In particular, for $0\leq \epsilon <1$, {$r_P\big(n,k=\Omega(n^{\epsilon^-})\big) = \cO\big(n^{\delta(\epsilon)^+}\big)$}, where $\delta(\epsilon) = \min_{{s:s > {\frac 1 {1-\epsilon}}} } \{\delta_s(\epsilon)\}$ and  $\delta_s(\epsilon)= 1-\frac {s(1-\epsilon)-1}  {2s(s-1)} $, and $r_P\big(n,k=\Theta(n^{\epsilon})\big) = o(n)$.
\end{theorem}

The analysis so far dealt with constructing $k$-PIR when $k=\Theta(n^\epsilon)$ and $0 \leq \epsilon<1$. Now we show how to use these results to construct $k$-PIR codes for $\epsilon \geq 1$. The idea is to concatenate a sufficient copies of $k'$-PIR codes, when $k'=\Omega(n^{1^-})$ such that each bit will have $k$ recovering sets.\vspace{-2ex}
\begin{theorem}\label{theorem_PIR_k>n}
For all $\epsilon \geq 1$ and $n$ sufficiently large, there exists a binary $k$-PIR code $[N,n,k]_2^P$, such that  $k=\Theta(n^{\epsilon})$ and $N=\cO(n^{\epsilon^+})$. 
\end{theorem}\vspace{-1ex}

\begin{figure}
\begin{tikzpicture}
  \begin{axis}[
  width=9cm,     
  axis y line=center,
  axis x line=left,
  scaled ticks=false,
  axis equal,
  grid=major,
  xmax=2,xmin=0,
  ymin=0,ymax=2,
  xlabel=$\epsilon$,ylabel=$\delta$,
  legend pos=south east,
  ]

	\addplot+[domain=0:1,restrict y to domain=0:1, thick,mark=none] {1-(3-1-x*3)/(2*3*(3-1))};
	\addlegendentry{$\delta_3(\epsilon)$}
	\addplot+[domain=0:1,restrict y to domain=0:1, thick,mark=none] {1-(5-1-x*5)/(2*5*(5-1))};
	\addlegendentry{$\delta_5(\epsilon)$}
	\addplot+[domain=0:1,restrict y to domain=0:1, thick,mark=none] {1-(7-1-x*7)/(2*7*(7-1))};
	\addlegendentry{$\delta_7(\epsilon)$}
	\addplot+[domain=0:1,restrict y to domain=0:1, thick,mark=none,color=black] {1-(20-1-x*20)/(2*20*(20-1))};
	\addlegendentry{$\delta_{20}(\epsilon)$}

	\addplot+[domain=1:3,thick] {x};
	\addlegendentry{$\epsilon\geq 1$}
	
	\addplot+[color=green,thick] coordinates { (0,0.5) (0.29,0.79) (0.5,0.79) (0.5,1) (1,1.5) (1.5,2)  };
	\addlegendentry{old results}
	
	\addplot+[color=red] coordinates { (0,0.5) (0.5,0.5) (1,1) (1.5,1.5) (2,2)  };
	\addlegendentry{lower bound}
  \end{axis}
\end{tikzpicture}\vspace{-3.5ex}
\caption{Asymptotic results for binary PIR codes}\label{fig:binaryPIRlong}
\end{figure}
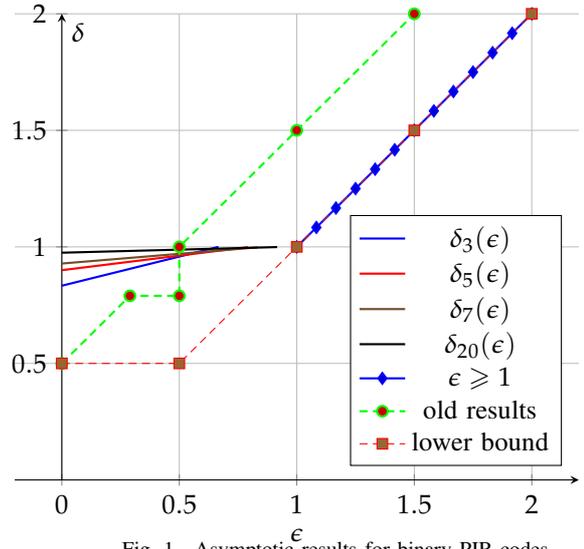

The length achieved by the PIR construction in Theorem~\ref{theorem_PIR_k>n} is nearly optimal. Recall that the length of $k$-PIR codes is $\Omega(k)$ since every non-trivial recovering set must contain at least one redundancy bit. 
Fig.~\ref{fig:binaryPIRlong} summarizes the results of binary PIR codes we achieved in this section together with the previous results. We plot the curves $\delta_s(\epsilon)$ for $s=3,5,9, 20$ from Theorem~\ref{PIR_binary_summary} as well as the results for $\epsilon\geq 1$ from Theorem~\ref{theorem_PIR_k>n}. The lower bound on the redundancy is given by $\min\{k,\sqrt{n}\}$.

\section{Batch codes from multiplicity codes}\label{sec:multi_batch}
It turns out that multiplicity codes can be also an excellent tool to construct batch codes. Unlike the PIR case, recovering different entries in the codeword will cause intersection in the corresponding lines, and thus intersecting recovering sets. In order to overcome this obstacle, we reduce the degree $d$ of the polynomials such that a fewer number of points is needed from every line. This will allow different lines to avoid points which are used by other lines. That way, every recovering set can "drop out" points which are used by other sets, resulting in disjoint recovering sets.\vspace{-1ex}

\begin{lemma}\label{lem:batchMC}
	For all $m,s,q,d,k$ such that $d\leq{m(q-km^{s-1}-2)}$ and $k\leq{\floor{\frac q {m}}^{s-1}}$, the code $\cC(m,d,s,q)$ is a $k$-batch code $[q^s,n,k]_Q^B$, where $n=\frac { { {d+s} \choose {s}  }}  {  { {s+m-1} \choose {s}  }  }$ and $Q= q^{{s+m-1} \choose {s}}$. 
\end{lemma}
\begin{IEEEproof}
The claim regarding the code dimension and field size can be proven similarly to PIR codes. Now we prove that every multiset request of size $k$ can be recovered. As we saw in the recovering procedure for PIR codes, every recovering set contains $m^{s-1}$ different lines. Since different lines can intersect on at most one point, and there are $k$ recovering sets, it suffices to prove that Step 1 in the recovering procedure can be completed even when $km^{s-1}$ points on the line are not used. But since the minimum distance of $\cC(m,d,s=1,q)$ equals $q-\frac d m>km^{s-1}+1$, it can be shown in a very similar way to PIR codes, that the polynomial $p_\bfv(\lambda)$ in Step 1 can be uniquely recovered, and thus also Step 2 can be completed.	
\end{IEEEproof}

Unlike the PIR case, it turns out that only the value $s = 2$ is useful for batch codes, thus getting the following theorem.\vspace{-2ex}

\begin{theorem}\label{theorem_batch_MC}
	For every $0<\alpha<0.5$ and $n$ sufficiently large, there exists a $k$-batch code $[N,n,k]_Q^B$ over $\F_Q$ with redundancy $r=N-n$ such that
	\begin{align*}
	k= \Theta(n^{0.5-\alpha}), 		r = \cO(n^{1-\frac {\alpha} 2 }),	    Q = n^{\Theta(n^\alpha)}.
	\end{align*}
	In particular, for $0<\epsilon<0.5$, it holds that $r_B\big(n,k=\Theta(n^\epsilon)\big) = \cO\big(n^{\delta(\epsilon)}\big)$, where $\delta(\epsilon) = \frac 3 4 + \frac \epsilon 2$.
\end{theorem}

As in the PIR case,  the last result can be extended for binary batch codes.\vspace{-2ex}
\begin{theorem}\label{theorem_batch_MC_binary}
	For every {$0<\alpha<{0.5}$} and $n$ sufficiently large, there exists a binary $k$-batch code $[N,n,k]^B$ with redundancy $r=N-n$ such that
	\begin{align*}
		{k= \Theta\big((n/ \log(n))^{0.5-\alpha}\big)}, 	r = \cO\big(n^{1-\frac \alpha 3 }(\log(n))^{\frac \alpha 3 }\big).
	\end{align*}
	In particular, for $0<\epsilon<0.5$, it holds that $r_B\big(n,k=\Omega(n^{\epsilon^-})\big) = \cO\big(n^{\delta(\epsilon)^+}\big)$, where $\delta(\epsilon) = \frac 5 6 + \frac \epsilon 3$, and $r_B\big(n,k=\Theta(n^{\epsilon})\big) = o(n)$. For $\epsilon \geq 0.5$ there exists a binary $k$-batch code $[N,n,k]_2^B$ of dimension $n$ such that $k=\Theta(n^{\epsilon})$ and $N=\cO(n^{0.5+\epsilon^+})$.
\end{theorem}

\section{Array Construction for PIR and Batch Codes}\label{sec:arrayPIR}

Our point of departure for this section is the subcube construction from~\cite{IKOS04} which was also used in~\cite{FVY15} to construct PIR codes. The idea of this construction is to position the information bits in a two-dimensional array, and add a simple parity bit for each row and each column. {Our approach here is to extend this construction by considering also diagonals with different slopes. As there are many different slopes, this can greatly increase the number of recovering sets}. However, we will have to guarantee that using the diagonals will still result with disjoint recovering sets. By a slight abuse of notation, in this section we let the set $[n]$ denote the set of integers $\{0,1,\ldots,n-1\}$. We use the notation $\langle x\rangle_m$ to denote the value of $(x\bmod m)$.\vspace{-2ex}

\begin{definition}\label{diag_def}
	Let $A$ be an $r\times p$ array, with indices $(i,j)\in[r]\times [p]$. For $s\in [p]$ we define the following set of sets $P_s(r,p) = \{D_{s,0},D_{s,1},\dots,D_{s,p-1}\},$ where for $t\in [p]$,
	\begin{equation*}
	\hspace{-0.7ex}D_{s,t}\hspace{-0.3ex}=\hspace{-0.3ex}\{(0,t),(1,\langle t+s\rangle_p),\ldots,(r-1,\langle t+(r-1)s\rangle_p) \}
	\end{equation*}
\end{definition}
The idea behind Definition~\ref{diag_def} is to fix a slope $s\in[p]$ and then define $p$ diagonal sets which are determined by the starting point on the first row and the slope. We use these sets in order to construct array codes, where every diagonal determines a parity bit for the bits on this diagonal.\vspace{-2ex}

\begin{cons}[Array Construction]\label{cons_diag}
	Let $r,p,n$ be positive integers such that $n=rp$, and $S\subseteq [p]$ a subset of size $k$. We define the encoder $E_{r,p,S}$, as a mapping $E_{r,p,S}:\{0,1\}^n\rightarrow \{0,1\}^{k\cdot p}$ as follows. We denote $S=\{s_0,s_1,\ldots,s_{k-1}\}$ where $0\leq s_0<s_1<\cdots <s_{k-1}\leq p-1$.The input vector $\bfx\in\{0,1\}^n$ is represented as an $r\times p$ array, that is $\bfx=(x_{i,j})_{(i,j)\in[r]\times[p]}$ and is encoded to the following $kp$ redundancy bits $\rho_{\ell,t}$, for $\ell\in [k]$, and $ t\in[p]$,
	$$\rho_{\ell,t} = \sum_{(i,j)\in D_{s_\ell,t}}x_{i,j}.$$ 
	Let $E_{r,p,S}(\bfx) = (\rho_{0,0},\ldots,\rho_{0,{p-1}},\ldots,\rho_{k-1,0},\ldots,\rho_{k-1,{p-1}})$, and the code $\cC(r,p,S)$ is defined to be
	$$\cC(r,p,S)   =  \{(\bfx,E_{r,p,S}(\bfx)) \ : \ \bfx\in\{0,1\}^n  \}.$$
\end{cons}

We first list several useful properties.  \vspace{-2ex}
\begin{lemma}\label{lem:partition}
	For all $r,p$, and $s\in [p]$ the set $P_s(r,p)$ is a partition of $[r]\times [p]$.
\end{lemma}
\begin{lemma}\label{lem:intersect}
	For all $r\leq p$ and $S\subseteq [p]$. If $p$ is prime, then for all ${s_1\neq{s_2}}\in{S}$ and $t_1,t_2\in [p]$, $|D_{s_1,t_1}\cap{D_{s_2,t_2}}|\leq 1$.
\end{lemma}

We only state here the result of this construction for PIR codes, as we focus here mainly on batch codes.\vspace{-2ex}
\begin{theorem}
	Let $n=p^2$, where $p$ is a prime number, and  $k\leq{\sqrt{n}}$. The code $\cC(r=p,p,S=[k])$ is a $k$-PIR code with redundancy $k\sqrt{n}$. In particular, for all $k\leq\sqrt{n}$, $r_P(n,k)=\cO(k\sqrt{n})$.
\end{theorem}

For batch codes, this construction can result with good batch codes as well as batch codes with restricted size for the recovering sets~\cite{ZS16}. Formally, a $k$-PIR code, $k$-batch code, in which the size of each recovering set is at most $r$ will be called an \emph{$(r,k)$-PIR code}, \emph{$(r,k)$-batch code}, respectively. 

The idea here is to choose the set $S$ in a way that for every bit, each of its recovering sets intersects with at most one recovering set of any other bit. This property for constructing batch codes from PIR codes was proved in~\cite{RSDG16} and is stated below.\vspace{-2ex}
\begin{lemma}\label{Batch_condition}
	Let $\cC$ be an $(r,k)$-PIR code. Assume that for every distinct indices $i,j\in[n]$, it holds that each recovering set of the $i$th bit intersects with at most one recovering set of the $j$th bit. Then, the code $\cC$ is an $(r,k)$-batch code.
\end{lemma}

The main challenge is to find sets $S$ that will generate recovering sets which satisfy the condition in Lemma~\ref{Batch_condition}. For that, we use the following definition. \vspace{-1ex}
\begin{definition}
	Let $r$ be a positive integer, and  $S$ be a set of non-negative integers. We say that the set $S$ does not contain an \textbf{{$r$-weighted arithmetic progression modulo $p$}} if there do \emph{not} exist $s_1,s_2,s_3\in{S}$ and $0<x,y<r-1$, where $x+y<r$, such that $xs_1+ys_2=(x+y)s_3 \bmod p $.
\end{definition}
Given this definition, we prove the following theorem.\vspace{-2ex}
\begin{theorem}\label{Batch_Cond}
	Let $r\leq p$ and $S\subseteq [p]$, $|S|=k$. If $p$ is prime, and $S$ does not contain an $r$-weighted arithmetic progression modulo $p$, then the code $\cC=\cC(r,p,S)$ is an $(r,k)$-batch code of dimension $rp$. \vspace{-1ex}
\end{theorem}

\begin{IEEEproof}
	Assume that $S=\{s_0,s_1,,\ldots,s_{k-1}\}$. One can verify using Lemma \ref{lem:partition} and \ref{lem:intersect} that for every $(i,j)\in[r]\times [p]$ the following sets 
	$$R_\ell^{(i,j)}=\{ \rho_{\ell,t_\ell} \}\cup{ \{x_{i',j'}\ : \ (i',j')\in D_{s_{\ell},t_\ell}\setminus \{(i,j)\} \} },\vspace{0ex}$$
	for $\ell \in[k]$ are $k$ mutually disjoint recovering sets for $x_{i,j}$, where $t_\ell \in[p]$ is chosen such that $(i,j)\in D_{s_{\ell},t_\ell}$. We denote $D(R_\ell^{(i,j)}) = D_{s_{\ell},t_\ell}$. Thus $\cC$ is $(r,k)$-PIR, and it remains to prove that $\cC$ satisfies the condition of Lemma~\ref{Batch_condition}. Assume in the contrary that there exist two bits $(i,j),(i',j')\in[r]\times [p]$ such that $(i,j)$ has a recovering set $R_{\ell_1}^{(i,j)}$ that intersects with two recovering sets $R^{(i',j')}_{\ell_1'},R^{(i',j')}_{\ell_2'}$ of $(i',j')$. Assume $b_1\in R_{\ell_1}^{(i,j)}\cap{R^{(i',j')}_{\ell_1'}}$ and  $b_2\in R_{\ell_1}^{(i,j)}\cap{R^{(i',j')}_{\ell_2'}}$ where $b_1,b_2$ are codeword entries. It can be verified that $b_1,b_2$ don't correspond to parity bits. Therefore, we denote $b_1=x_{i_1,j_1}$, $b_2=x_{i_2,j_2}$, for $(i_1,j_1),(i_2,j_2)\in[r]\times [p]$. Denote $D(R_{\ell_1}^{(i,j)})=D_{s'_1,t_1},D(R^{(i',j')}_{\ell_1'})=D_{s'_2,t_2},D(R^{(i',j')}_{\ell_2'})=D_{s'_3,t_3}$ for $s'_1,s'_2,s'_3\in{S}$ and $t_1,t_2,t_3\in[p]$.  Thus we get that
	$ (i_1,j_1),(i_2,j_2)\in{D_{s'_1,t_1}}, (i_1,j_1),(i',j')\in{D_{s'_2,t_2}}$, and $ (i_2,j_2),(i',j')\in{D_{s'_3,t_3}}$. From Lemma \ref{lem:partition} and \ref{lem:intersect} we deduce that $s'_1\neq s'_2 \neq s'_3$ and $i'\neq i_1 \neq i_2$. Assume w.l.o.g $i_1<i_2<i'$. It follows that:
	\begin{align*}
	j_1=& \langle t_1 + i_1s'_1\rangle_p, \quad j_2 = \langle t_1 + i_2s'_1\rangle _p\\
	j_1 =&  \langle t_2 + i_1s'_2\rangle_p, \quad j'=  \langle t_2 + i's'_2\rangle_p \\
	j_2=&  \langle t_3 + i_2s'_3\rangle_p, \quad j' = \langle t_3 + i's'_3\rangle_p
	\end{align*}
	This implies that
	$\langle(i_2-i_1)s'_1 + (i'-i_2)s'_3\rangle_p = \langle(i'-i_1)s'_2\rangle_p$, which is a contradiction since $S$ does not contain an $r$-weighted arithmetic progression modulo $p$.
\end{IEEEproof}
In order to complete the construction of batch codes, we are left with the problem of finding large sets $S$ which satisfy the condition in Theorem~\ref{Batch_Cond}. That is, given $r$ and $p$, our goal is to find the largest such a set $S$. A simple greedy algorithm can give the following result.\vspace{-2ex}
\begin{theorem}\label{G2}
Let $r,p$ be positive integers, such that $p$ is prime. Then there exists a set $S$ with no $r$-weighted arithmetic progression modulo $p$ of size at least $k$, where $k$ is the largest integer such that $p>2k^2r^2$.
\end{theorem}

The following theorem follows from these observations.\vspace{-1ex}
\begin{theorem}\label{rBatch_theorem}
	For every $r,k$, let $n=rp$, where $p$ is the smallest prime number such that $2k^2r^2<p$. Then, there exists an $(r,k)$-batch code of dimension $n$ and rate $\frac r {r+k}$. In particular, the redundancy of the code equals $kp$. \vspace{-1ex}
\end{theorem}
According to Theorem~\ref{rBatch_theorem} we are now at a point to construct $k$-batch codes with good redundancy. \vspace{-1ex}
\begin{corollary}\label{array_1}
	For any $n$ and $k$ such that $k=o(\sqrt{n})$, there exists a $k$-batch code of dimension $n$ and redundancy $\cO(n^{\frac 2 3}k^{\frac 5 3})$. In particular, for $0<\epsilon < 1/2$, $r_B(n,n^\epsilon) =\cO(n^{2/3+5\epsilon/3})$. \vspace{-1ex}
\end{corollary}
\begin{IEEEproof}
	For $n$ and $k$, let us choose $r= \lceil{n^{\frac 1 3}}/ {k^{\frac 2 3}}\rceil$, and $p$ is the smallest prime number such that $2k^2r^2<p$. Then, according to Theorem~\ref{rBatch_theorem}, there exists an $(r,k)$-batch code of dimension $pr>n$ and redundancy $kp$. That is, the redundancy satisfies $kp = \Theta(k^3r^2) = \Theta(n^{\frac 2 3}k^{\frac 5 3}).$ The second statement in the corollary is established for $k=n^\epsilon$ in the last equation.
\end{IEEEproof}

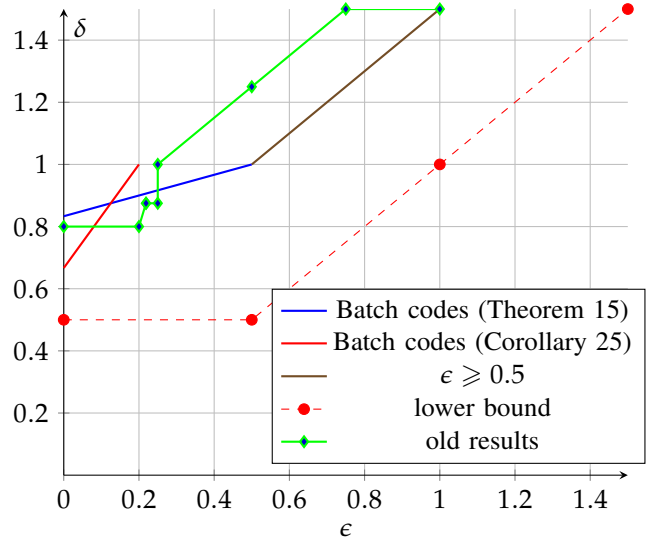
\begin{figure}
\begin{tikzpicture}
\begin{axis}[
width=9cm,             
axis y line=center,
axis x line=left,
grid=major,
xmax=1.5,xmin=0,
ymin=0,ymax=1.5,
xlabel=$\epsilon$,ylabel=$\delta$,
legend style={at={(0.7,0.4)},anchor=north},
]

\addplot+[domain=0:1/2,thick,mark=none] {5/6+x/(3)};
\addlegendentry{Batch codes (Theorem~\ref{theorem_batch_MC_binary})}

\addplot+[domain=0:0.2,thick,mark=none] {2/3+x/(3/5)};
\addlegendentry{Batch codes (Corollary~\ref{array_1})}

\addplot+[domain=0.5:2,thick,mark=none] {1/2+x};
\addlegendentry{$\epsilon\geq 0.5$}

\addplot+[color=red,dashed,mark=*,mark options={solid}] coordinates { (0,0.5) (0.5,0.5) (1,1) (1.5,1.5) (2,2)  };	\addlegendentry{lower bound}

\addplot+[color=green,thick] coordinates { (0,0.8) (0.2,0.8) (7/32,28/32) (7/32,7/8) (1/4,7/8) (1/4,1) (1/2,5/4) (3/4,1.5) (1,1.5) (1.5,2)  };
\addlegendentry{old results}

\end{axis}
\end{tikzpicture}\vspace{-2.5ex}
\caption{Asymptotic results for binary batch codes}\label{fig:batch}
\end{figure}

Let us denote $r_B(k=n^\epsilon) = \cO(n^\delta).$
In Fig.~\ref{fig:batch} we plot the results on the asymptotic behavior of the redundancy of batch codes. These plots are received from Corollary~\ref{array_1} in this section and Theorem~\ref{theorem_batch_MC_binary} from Section~\ref{sec:multi_batch}. Note that the array construction improves the redundancy only for $\epsilon < 0.0755$.

Lastly, we report on two more results that can be derived using the Array Construction. Note that the second result improves upon the one from~\cite{VY16}, which states that $r_B(n,5)=\cO(\sqrt{n}\log(n)$. 
\vspace{-1ex}
\begin{theorem}\label{rbatch_3}
For every $0<\alpha<1$, $k=\cO(n^\alpha)$, and fixed $r\geq 3$, there exists an $(r,k)$-batch code with rate $\frac r {r+k}$. \vspace{-1ex}
\end{theorem} 
\begin{theorem}
	Let $n=p^2$ where $p$ is a prime number. The code $\cC$, that extends $\cC(r=p,p,S=[5])$ by adding a global parity bit, is a $5$-batch code with redundancy $5p+1=\Theta(\sqrt{n})$, and therefore $r_B(n,5)=\Theta(\sqrt{n})$. 
\end{theorem}

\end{document}